\documentclass[conference]{IEEEtran}
\IEEEoverridecommandlockouts

\usepackage{cite}
\usepackage{amsmath,amssymb,amsfonts}
\usepackage{algorithmic}
\usepackage{graphicx}
\usepackage{textcomp}
\usepackage{xcolor}
\usepackage{hyperref}
\usepackage{booktabs} 
\usepackage{algorithm}
\usepackage{algorithmic}

\def\BibTeX{{\rm B\kern-.05em{\sc i\kern-.025em b}\kern-.08em
    T\kern-.1667em\lower.7ex\hbox{E}\kern-.125emX}}
\begin{document}

\title{CryptoGAT: Are Time Series Models Effective for Cryptocurrency Forecasting?\\
}

\author{
\IEEEauthorblockN{1\textsuperscript{st} Yu Peng}
\IEEEauthorblockA{\textit{School of Computer Science} \\
\textit{University of Sydney}\\
Sydney, Australia \\
ypen0278@uni.sydney.edu.au}
\and
\IEEEauthorblockN{2\textsuperscript{nd} Matloob Khushi}
\IEEEauthorblockA{\textit{Department of Computer Science} \\
\textit{Brunel University London}\\
Uxbridge, United Kingdom \\
Matloob.Khushi@brunel.ac.uk}
\and
\IEEEauthorblockN{3\textsuperscript{rd} Josiah Poon}
\IEEEauthorblockA{\textit{School of Computer Science} \\
\textit{University of Sydney}\\
Sydney, Australia \\
josiah.poon@sydney.edu.au}

}

\maketitle

\begin{abstract}
Cryptocurrency price prediction is a significant challenge in quantitative investment. In recent years, time series models have made significant progress in financial forecasting tasks, especially in the stock market. Despite the growing performance over the past few years, we question the validity of this line of research in cryptocurrency prediction. Specifically, time series models (e.g., LSTM, GRU, and Transformers) are effective at extracting temporal relationships in stock market data. However, in pure price-based cryptocurrency prediction, facing data with extreme volatility and wild swings, time series models have difficulty learning effective information. To validate our claim, we propose CryptoGAT, a lightweight Graph Attention Network that recasts cryptocurrency pure price prediction as a cross-asset graph problem rather than a temporal modeling task. Extensive experiments on real cryptocurrency benchmarks demonstrate that our proposed CryptoGAT outperforms various state-of-the-art forecasting methods with a notable margin. Moreover, we conduct comprehensive empirical studies to explore the fundamental differences exposed by time series models in stock and cryptocurrency prediction: differences in predictability of the signal and cross-asset dependencies. This finding opens up new research directions for the cryptocurrency pure price prediction task and inspires further graph-based exploration in the field. The source code is available at \url{https://github.com/FanBroWell/CryptoGAT}.
\end{abstract}


\section{Introduction}
Financial asset price forecasting is an important task in the domain of quantitative investment. Among the universe of financial assets, cryptocurrencies hold uniquely disruptive potential, primarily underpinned by decentralized architecture. The 2008 Bitcoin whitepaper \cite{nakamoto2008} brought a decentralized electronic cash system and Bitcoin, the first cryptocurrency, into the mainstream consciousness. In recent years, the impact of cryptocurrencies on the world has become increasingly significant \cite{hardle2020understanding}. The decentralized architecture not only has profound implications for reshaping future social structures \cite{lumineau2021blockchain} and driving economic reforms \cite{corbet2019cryptocurrencies}, but also introduces significant changes to the field of medical and healthcare \cite{agbo2019blockchain}. Cryptocurrencies are now widely recognized as a major financial asset class, with a total market capitalization \cite{ReutersCrypto2025} approximating \$4 trillion by May 2025. Despite the growing importance of cryptocurrencies, academic research remains mainly focused on traditional stock markets, leaving the cryptocurrency markets pitifully underexplored, especially in terms of pure price prediction.

The focus of research in the financial field has been on developing more accurate stock investment prediction models for a long time. In the initial phase, researchers primarily employed traditional machine learning to identify patterns in financial data \cite{strader2020machine, gu2020empirical}. Representative techniques of this period include decision trees \cite{kamble2017short, nugroho2014decision}, support vector machines (SVM) \cite{xie2013semantic}, and the k-nearest neighbor algorithm (KNN) \cite{alkhatib2013stock}. With the rise of deep learning, academic research has shifted its focus to designing more complex neural network architectures \cite{sezer2020financial, pang2020innovative}. Specifically, existing research generally captures the predictability of the stock market from three dimensions: technical indicators correlations \cite{brock1992simple}, temporal correlations, and asset correlations. However, unlike traditional assets (such as stocks) with clear valuation bases and constraints imposed by regulation and trading hours, the cryptocurrency market's uninterrupted global trading mechanism and extremely unstable liquidity structure make it an \emph{atypical} financial environment \cite{liu2021risks}. Notable features include:

\begin{itemize}
    \item \textbf{24/7 Operation:} Operates continuously globally, without a unified trading session or market closure schedule set by a central authority.
    \item \textbf{Lack of Intrinsic Valuation:} Without the support of tangible assets and fundamental disclosures.
    \item \textbf{Extreme Volatility:} Driven by both market narratives and liquidity impulses, highly sensitive to sentiment feedback.
    \item \textbf{Irregular Liquidity:} Many tokens suffer from inconsistent liquidity, exacerbating price impact and risk exposure.
\end{itemize}

\noindent An important observation is that the fundamental reason hindering cryptocurrency prediction research lies in the \textbf{bias in problem definition}. Due to the significant time dependence of financial assets, they have long been treated as multivariate time series prediction problems. The main working power of the Time Series Model is from its powerful ability to extract time sequence dependencies. Essentially, they capture the time patterns of evolution. However, cryptocurrencies are not inherently driven by time patterns, but rather constitute a \textbf{Graph Network}. Although the data is continuous in time, the driving force behind price fluctuations often does not originate from its ``time-series," but rather from ``other spatial assets". Consequently, we pose the following intriguing question: \textbf{Are Time Series Effective for Cryptocurrency Forecasting?}

Furthermore, we note that high-quality, clean price-only datasets for cryptocurrency prediction are inherently scarce \cite{alexander2020critical}. Unlike equities, the crypto market is populated by thousands of tokens with extremely poor liquidity, brief lifespans, and fragmented exchange coverage \cite{john2024cryptocurrency, dimpfl2021nothing}. This \textbf{Data Scarcity} is itself a core challenge of cryptocurrency prediction. This constraint has driven the field research toward incorporating external and multimodal signals (such as on-chain data \cite{feng2025cryptomixer} and sentiment features \cite{gurgul2025deep}) to compensate for the limited predictive power of price data alone. 

Our work takes the opposite approach: rather than augmenting external data sources, we recast the problem itself as a graph problem, demonstrating strong performance under pure price-based constraints. Therefore, we introduce a set of simple yet effective models named \textbf{CryptoGAT} for comparison. CryptoGAT uses an attention mechanism architecture to process graph-structured data, allowing each node to focus on its neighbors to compute hidden representations and assign different weights. We evaluate our approach on a real-world cryptocurrency dataset collected from the Binance exchange, the world’s largest cryptocurrency trading platform by volume. Remarkably, our results show that CryptoGAT outperforms existing complex Time-series models in all cases, and often by a large margin. Finally, we conduct comprehensive empirical studies on existing solutions and datasets. To sum up, the contributions of this work include:

\begin{itemize}
    \item To the best of our knowledge, this is the first work to challenge the effectiveness of time series models for the cryptocurrency pure price forecasting task by empirical analysis.
    \item We identify and recast the cryptocurrency pure price prediction task as a graph problem, which improved prediction accuracy. The new direction we propose inspires further graph-based exploration in the field.
    \item To validate our claims, we introduce a set of simple yet effective models named CryptoGAT, and compare them with existing SOTA solutions. Through rigorous, reproducible experiments, we show that our work achieves strong performance. This simple graph structure itself also underscores the importance of redefining the problem. 
    \item We conduct comprehensive empirical studies on various aspects of existing solutions, revealing the limitations of time series models and the unique characteristics of cryptocurrency data. Our findings would benefit future research in this area.
\end{itemize}

\noindent With the above, we find that the fundamental reason hindering cryptocurrency prediction research lies in the bias in problem definition. To solve this problem, we provide a new problem formulation that revisits an underlying bias present in existing definitions within this research area. At the same time, our proposed CryptoGAT achieves better prediction results than existing works and highlights the potential of developing a graph framework for future research on cryptocurrency prediction tasks.

\section{Related Work}

In this section, we review the related work from the literature of Time Series Forecasting and Financial Price Forecasting.

\textbf{Time Series Forecasting.}  Time series forecasting plays a critical role in many fields, such as economics, solar energy, traffic, weather, electricity, and finance. Early deep learning used recurrent neural networks (RNNs) and long short-term memory (LSTM) networks \cite{hochreiter1997long} to capture temporal dependencies. With the successful application of the Transformer architecture in natural language processing, researchers have applied attention mechanisms to time series prediction, resulting in models such as PatchTST \cite{nie2023time} and iTransformer \cite{liu2024itransformer}. These models effectively model temporal dependencies through self-attention mechanisms. Therefore, in our experiments, we selected two of the most representative Transformer-based models. PatchTST \cite{nie2023time} includes a special design: patching. Patch enhances the identification of local information. Each input time series is divided into different patches. This approach reduces the length of the input sequence and improves training efficiency. iTransformer \cite{liu2024itransformer} reverses the dimensionality processing logic of Transformer, treating each time step as a whole feature vector and applying a self-attention mechanism on the feature dimension rather than the time dimension. These models represent state-of-the-art Transformer architectures in time series forecasting.

Beyond pure temporal modeling, spatio-temporal approaches have emerged as an important direction that jointly captures temporal and spatial dependencies. Temporal Graph Networks (TGN) \cite{rossi2020temporal} uses graphs to propagate temporal information. Graphormer \cite{ying2021transformers} incorporates graph structures into the Transformer's self-attention mechanism. MASTER \cite{li2024master} uses market information to guide Transformer attention along the time dimension. These representative works have demonstrated powerful performance in areas such as traffic forecasting and stock markets. However, these approaches focus more on the complexity of the time dimension, with graphs serving only as aids to the time model. We demonstrate, by introducing a simple graph attention model, that in the cryptocurrency market, cross-asset graph structure is more important than complex time modeling.

\textbf{Financial Price Forecasting.}  The evolution of price forecasting has long been based on the analysis of historical price-volume indicators. In the initial stages of this domain, research predominantly employed conventional mathematical algorithms to process numerical features derived from technical analysis standards \cite{piccolo1990distance, wang1996stock, tseng2002combining}. With the development of deep neural networks, early studies employed recurrent neural networks (RNNs) \cite{nelson2017stock, qin2017dual} and CNNs \cite{tsantekidis2017forecasting} to forecast short-term trends. To further capture fine-grained features, subsequent research has introduced advanced mechanisms such as self-attention \cite{li2018stock, ding2020hierarchical} and gated causal convolutions \cite{wang2021hierarchical}. Recent research has focused on the importance of inter-asset relationships. For example, RSR \cite{feng2019temporal} combines temporal encoding with relational embeddings through a graph convolution mechanism. To better capture the dependencies between assets, ESTIMATE \cite{huynh2023efficient} combines hypergraphs with temporal generative filters to enable the modeling of non-pairwise market correlations. The latest state-of-the-art model, StockMixer \cite{fan2024stockmixer}, has been designed with three mixing modules to effectively model technical metrics, time dependencies, and space dependencies. Although StockMixer performs well on stock prediction tasks, its effectiveness does not transfer to cryptocurrency forecasting.

\section{Methodology}
\label{sec:method}

\subsection{Problem Definition}

Following the setup of existing works \cite{huynh2023efficient, feng2019enhancing}, we adapt the stock price forecasting framework to cryptocurrency markets. We input historical price patterns with multiple indicators (Open, High, Low, Close prices, and Volume) and output the closing price of the next day to calculate the 1-day return ratio.

Given a cryptocurrency market composed of $N$ cryptocurrencies $X = \{X_1, X_2, \ldots, X_N\}$, each cryptocurrency is a two-dimensional matrix $X_i \in \mathbb{R}^{T \times F}$, where $T$ is the lookback window length and $F$ is the number of input features. Our target is to predict the closing price $p_i^t$ on trading day $t$ and calculate the 1-day 
return ratio:

\begin{equation}
r_i^t = \frac{p_i^t - p_i^{t-1}}{p_i^{t-1}}
\label{eq:return}
\end{equation}

\subsection{Time Series Models}

\subsubsection{Theoretical Foundation}

Time series models assume that future asset prices can be inferred from historical sequences. A canonical example is the autoregressive (AR) model, which expresses the current value as a weighted combination of its past observations:

\begin{equation}
y_t = \phi_1 y_{t-1} + \phi_2 y_{t-2} + \cdots + \phi_z y_{t-z} + \epsilon_t \approx f(\mathbf{X}_{t-1})
\label{eq:ar_model}
\end{equation}

\noindent Where $y_t$ represents the observation at time $t$, $\phi_i$ are the parameters learned by the model, $z$ is the lag order, and $\epsilon_t$ denotes the white noise, $f(X_{t-1})$ is a function representation.

Time series models \cite{nelson2017stock, qin2017dual, tsantekidis2017forecasting, li2018stock, ding2020hierarchical, wang2021hierarchical} have long performed well in the stock market. Macroeconomic trends and financial reports in the stock market are stronger signals than random noise, and this high signal-to-noise ratio makes it easier for models to capture effective information and achieve better predictive performance. For example, the latest state-of-the-art (SOTA) model, StockMixer, has achieved excellent results with high IC and high Sharpe ratios on multiple large stock datasets.

\subsubsection{StockMixer as Representative Model}

\begin{figure}[h]
  \centering
  \includegraphics[width=\dimexpr\linewidth-2em\relax]{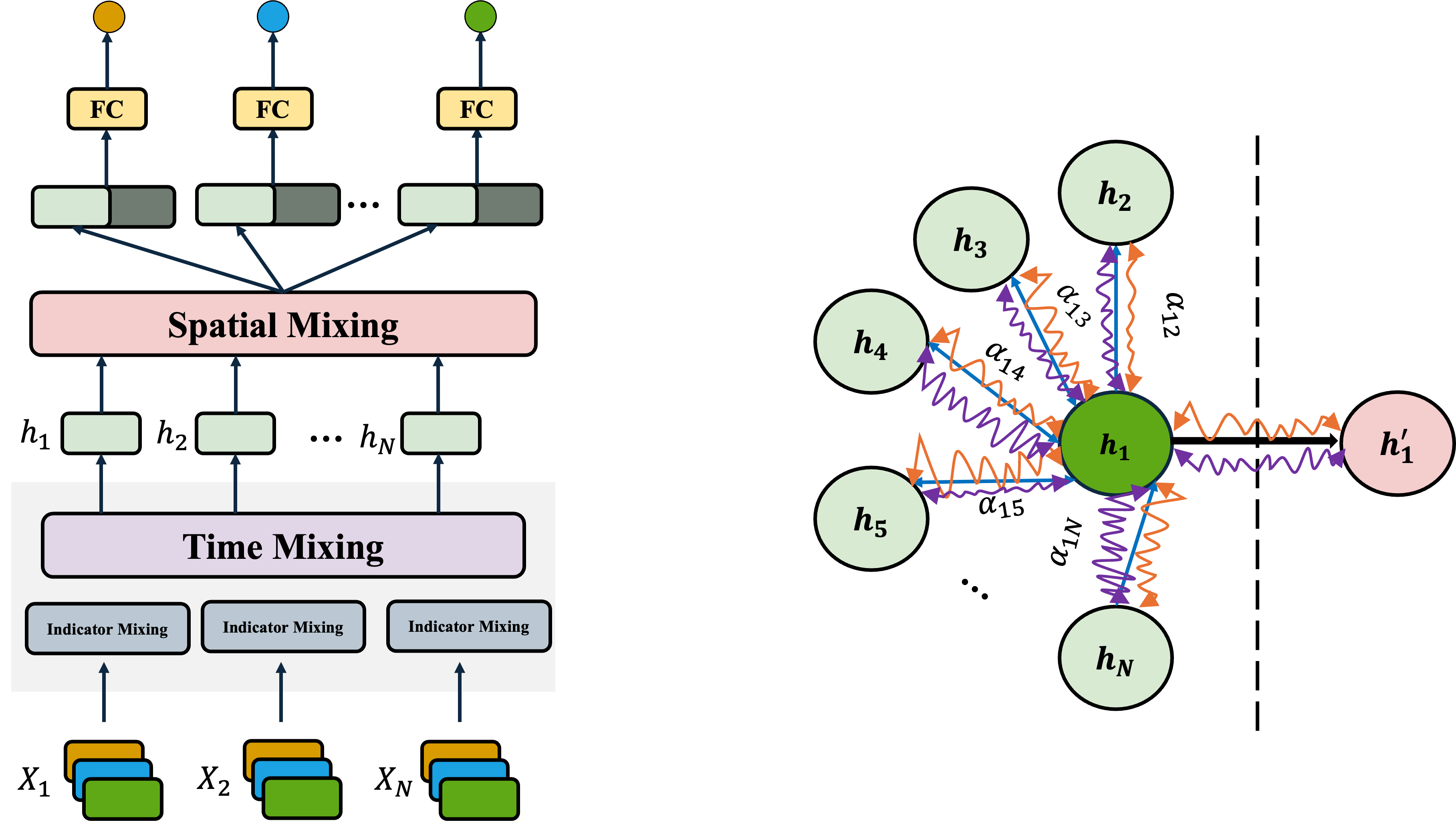}
  \caption{\textmd{\textbf{Left:} Overview of the StockMixer. 
           \textbf{Right:} The multi-head attention mechanism of the CryptoGAT model, 
           where $h_1$ obtains $\mathbf{h}_1'$ by calculating the weights of its neighbors.}}
  \label{fig:overview}
\end{figure}

We select StockMixer \cite{fan2024stockmixer} as our representative time series model for several reasons. First, StockMixer achieves superior performance on stock market benchmarks (NASDAQ Sharpe: 1.465, IC: 0.043), establishing it as a strong baseline for time series-based financial forecasting. Then, as shown in Figure~\ref{fig:overview}, StockMixer primarily contains three components: Indicator mixing, Time mixing, and Stock mixing. The indicator and time mixing modules extract assets' indicator information and time information. The Stock mixing module extracts the relationships between the assets. Finally, they combine these three representations to predict the closing price. This simple structure, which includes both time and spatial modules, helps us to conduct comparative experiments and verify the contributions of the modules.

\subsection{A Simple Yet Effective Baseline}

In existing financial time-series forecasting solutions, all the compared baselines (e.g., LSTM, GRU, Transformer) model each cryptocurrency independently, which ignores the strong cross-asset correlations in cryptocurrency markets. We proposed that cryptocurrency data is essentially a graph structure, altering the mathematical basis of cryptocurrency prediction tasks. Building upon the time series mathematical theory in formula 2, a new mathematical theory for cryptocurrency prediction tasks is recast, as shown in the following formula:

\begin{equation}
y_{t} \approx g(\mathbf{G}, f(\mathbf{X}_{t-1}))
\end{equation}

\noindent where $\mathbf{G}$ is graph structure. This implies that graph models based on asset relationships will perform better in cryptocurrency prediction. 

To validate this hypothesis, we present a simple yet effective graph-based model via a graph attention mechanism, named CryptoGAT, as a baseline for comparison. The model’s graph-constructed workflow is outlined (Algorithm~\ref{alg:graph}). The model constructs a graph structure by calculating the Pearson correlation coefficient. Based on a predefined graph, CryptoGAT learns the attention coefficients between assets, thereby identifying highly correlated assets and assigning them higher weights. Ultimately, by aggregating this key information, the model optimizes its predictive performance for future trends. The mathematical expression is:

\begin{equation}
\mathbf{h}_i' = \sigma\left(\sum_{j \in \mathcal{N}_i} \alpha_{ij} \mathbf{W}\mathbf{h}_j\right)
\end{equation}

\noindent where $\alpha_{ij}$ is the attention coefficient between cryptocurrency $i$ and $j$, $\mathcal{N}_i$ represents the set of cryptocurrencies correlated with cryptocurrency $i$. $\sigma$ is an activation function, computed as:

\begin{equation}
\alpha_{ij} = \frac{\exp(\text{LeakyReLU}(\mathbf{a}^T [\mathbf{W}\mathbf{h}_i \| \mathbf{W}\mathbf{h}_j]))}{\sum_{k \in \mathcal{N}_i} \exp(\text{LeakyReLU}(\mathbf{a}^T [\mathbf{W}\mathbf{h}_i \| \mathbf{W}\mathbf{h}_k]))}
\end{equation}

\noindent As shown in Figure~\ref{fig:overview}, $\mathbf{h}_i$ and $\mathbf{h}_j$ denote the feature representations of cryptocurrency $i$ and $j$, respectively. $\mathbf{W} \in \mathbb{R}^{F' \times F}$ is a learnable linear transformation matrix, $\mathbf{a} \in \mathbb{R}^{2F'}$ is a learnable attention vector, $\|$ denotes concatenation. 

\begin{algorithm}[t]
\caption{Correlation-Based Graph Construction}
\label{alg:graph}
\begin{algorithmic}[1]
\REQUIRE Price matrix $P \in \mathbb{R}^{N \times T_{\text{train}}}$, threshold $\tau$
\ENSURE Adjacency matrix $A \in \mathbb{R}^{N \times N}$

\STATE \textbf{Initialize:} $A \leftarrow \mathbf{0}^{N \times N}$
\FOR{$i = 1$ to $N$}
    \FOR{$j = i$ to $N$}
        \STATE $C_{ij} \leftarrow \text{corr}(P_i, P_j)$ \hfill $\triangleright$ Pearson correlation
        \IF{$|C_{ij}| > \tau$}
            \STATE $A_{ij}, A_{ji} \leftarrow |C_{ij}|$ \hfill $\triangleright$ Weighted undirected edge
        \ENDIF
    \ENDFOR
    \STATE $A_{ii} \leftarrow 1$ \hfill $\triangleright$ Self-loop
\ENDFOR
\RETURN $A$
\end{algorithmic}
\end{algorithm}

Note that CryptoGAT is a simple graph-based model. To handle cryptocurrency markets with different characteristics, we further introduce a variant, named FGAT:

\begin{itemize}
    
    \item \textbf{FGAT}: FeatureMixer Graph Attention Network (FGAT) combines the Indicator mixing mechanism with the graph attention architecture of GAT. First, it extracts more dimensional information from the raw data. Then, this richer, multi-dimensional information is fed into GAT, which uses graph attention to learn and dynamically assign weights between assets. This simple FGAT-based structure achieves remarkable predictive performance in cryptocurrency prediction.

\end{itemize}

\begin{table*}[t]
    \centering
    \caption{Comparison of different methods. \textmd{Better performance is reflected by higher metric values. \textbf{Bold} indicates best results, \underline{underline} indicates second best.}}
    \label{tab:main_results}
    \setlength{\tabcolsep}{4pt}
    \renewcommand{\arraystretch}{1.2}
    \large 
    \begin{tabular*}{\dimexpr\textwidth-2em\relax}{@{\extracolsep{\fill}}cc c ccccc@{}}
        \toprule
        \textbf{Model Type} & \textbf{Method} & \textbf{Parameters} & \textbf{IC} & \textbf{ICIR} & \textbf{Ann.SR} & \textbf{Prec@10} & \textbf{Inference time(ms)} \\
        
        \midrule
        \textbf{RNN}
                    & LSTM              & 51,521   & 0.011    & 0.046    & 2.299  & 0.496 & 0.252 \\
                    & ALSTM             & 55,746   & 0.009    & 0.043    & 1.162  & 0.503 & 0.276 \\
                    & GRU               & 38,657   & 0.017  & 0.072  & 1.529  & 0.495 & 0.245 \\
        \midrule
        \textbf{Transformer}
                    & PatchTST     & 266,369  & -0.004 & -0.003 & -0.256 & 0.501  & 0.410 \\
                    & iTransformer    & 13,764  & -0.007    &  -0.052   & -0.370  & 0.506 & 0.494 \\
        \midrule
        \textbf{GNN}
                    & RSR    & 76,481  & 0.015 & 0.079 & 1.544  & 0.499 & 0.449 \\
                    & ESTIMATE    & 332,259  & -0.002    & -0.025    & 1.018  & 0.496 & 7.899 \\
        \midrule
        \textbf{Spatio-Temp}      
                        & TGN            & 72,065 & 0.026 & 0.101  & 2.210  & 0.506 & 0.903 \\
                        & Graphormer           & 109,929 & -0.007 & -0.058  & -0.129  & 0.494 & 0.992 \\
                        & MASTER           & 142,593 & 0.027 & 0.114  & 2.227  & \textbf{0.513} & 1.382 \\
        \midrule
        \midrule
        \textbf{SOTA}
                    & StockMixer       & 2,327  & 0.011 & 0.113 & 0.356  & \underline{0.511} & 2.415 \\
        \textbf{CryptoGAT}
                    & \textbf{GAT}$^\dagger$ & 47,105 & \underline{0.037} & \underline{0.138} & \underline{3.128} & 0.504 & 0.627 \\
                    & \textbf{FGAT}$^\dagger$  & 52,865  & \textbf{0.047}    & \textbf{0.168}    & \textbf{3.892}  & 0.507 & 0.792 \\
        \bottomrule
        \multicolumn{8}{l}{\scriptsize  $^\dagger$Our proposed method; Spatio-Temp = Spatio-Temporal models.} \\
    \end{tabular*}
\end{table*}

\section{Experiments}
\label{sec:exp}

\subsection{Experimental Setup}

\textbf{Datasets.}  We evaluate our approach on a real-world cryptocurrency dataset collected from the Binance exchange, the world's largest cryptocurrency trading platform by volume. First, we selected the top 1000 cryptocurrencies by market capitalization. To address the common problems of cryptocurrencies, such as extremely poor liquidity and short lifespan \cite{alexander2020critical}, we rigorously filtered these 1000 cryptocurrencies: (1) market capitalization exceeding \$200M, (2) USDT-denominated trading pairs only, (3) continuous daily data with no missing values. After applying these filters, fewer than 20 of these tokens had continuous trading records prior to 2019. This filtering outcome itself illustrates the fundamental data scarcity challenge in cryptocurrency research.

Balancing these considerations, we selected daily data for 66 major cryptocurrencies between 15/04/2023 and 08/01/2026. Furthermore, we aligned all cryptocurrency data according to a common trading date. After time alignment, the data had 999 trading days, which covers multiple distinct markets from two consolidation-to-bull cycles, separated by the 2025 bear crash, subsequent recovery, and crash again. The captured raw data contains five dimensions: Opening price, Highest price, Lowest price, Closing price, and Volume, abbreviated as OHLCV data. To improve file reading speed and data stability, we processed the raw data, normalizing the price data relative to the previous day's closing price and the volume relative to its 5-day moving average. We sort the data in chronological order and split the dataset into training, validation, and testing sets with a ratio of 6:2:2. This strategy ensures that there is no overlap between the training and testing sets and avoids future information from interfering with the prediction of past behavior.

\textbf{Implementation Details.}  We implement our model on PyTorch and evaluate on an RTX 4090 24GB GPU and an NVIDIA A100 40GB GPU. Each experiment was repeated 3 times, and the average performance was reported. For fair comparison, all samples are generated by moving a 30-day lookback window along trading days. The loss factor $\alpha$ is 0.1, dropout is 0, and the learning rate is $2.5\mathrm{e}{-4}$.

\textbf{Metrics.}  Existing literature often adopts heterogeneous evaluation protocols, hindering a fair benchmarking of different methodologies. To ensure a rigorous and comprehensive comparison, our evaluation is consistent with StockMixer \cite{fan2024stockmixer} for four robust metrics, covering rank-based, accuracy-based, and return-based dimensions. \textit{The Information Coefficient} (IC) quantifies how close the predicted value is to the actual value by calculating the average Pearson correlation coefficient. \textit{The Information Coefficient Ratio} (ICIR) is the ratio of IC to its standard deviation, calculated by dividing by the time standard deviation of IC. ICIR assesses the stability and consistency of a model; a higher ICIR indicates more stable predictive performance. \textit{Precision@N} evaluates the hit rate within the top-$N$ predictions. For instance, when N is 10, and the labels of 5 among these top 10 predictions are positive, then the Precision@10 is 50\%. To balance profitability against risk, we utilize the \textit{Annualized Sharpe Ratio} (SR), which measures the average return per unit of volatility in relation to the risk-free rate. It is formulated as: $SR = \frac{R_t - R_f}{\theta}$, where $R_t$ denotes the portfolio return, $R_f$ is the risk-free rate, and $\theta$ represents the standard deviation of returns. All models use the trading protocol (Algorithm~\ref{alg:trading}), where $\sqrt{365}$ 
reflects the 24/7 nature of cryptocurrency markets.

\begin{algorithm}[t]
\caption{Trading Protocol}
\label{alg:trading}
\begin{algorithmic}[1]
\REQUIRE Predicted returns $\{\hat{r}_{i,t}\}_{i=1}^{N}$ for each day $t$, $K{=}5$
\ENSURE Annualized Sharpe Ratio $SR$
\FOR{each trading day $t = 1, \dots, T$}
    \STATE Rank assets by $\hat{r}_{i,t}$ in descending order
    \STATE Select top-$K$ assets; assign equal weight $w = 1/K$
    \STATE $r_{p,t} \leftarrow \frac{1}{K}\sum_{i \in \text{Top-}K} r_{i,t}^{\text{actual}}$
\ENDFOR
\STATE $Ann.SR \leftarrow \dfrac{\text{mean}(r_{p,1}, \dots, r_{p,T})}{\text{std}(r_{p,1}, \dots, r_{p,T})} \times \sqrt{365}$
\RETURN $Ann.SR$
\end{algorithmic}
\end{algorithm}

\textbf{Baselines.}  We compare our model with several state-of-the-art baselines, including LSTM \cite{hochreiter1997long}, ALSTM \cite{feng2019enhancing}, GRU \cite{cho2014learning}, RSR \cite{feng2019temporal}, PatchTST \cite{nie2023time}, iTransformer \cite{liu2024itransformer}, ESTIMATE \cite{huynh2023efficient}, StockMixer \cite{fan2024stockmixer}, TGN \cite{rossi2020temporal}, Graphormer \cite{ying2021transformers}, MASTER \cite{li2024master}. Detailed descriptions are introduced in Appendix A.

\subsection{Overall Comparison}

Table~\ref{tab:main_results} shows the performance of all the comparison methods. To ensure fairness, all baseline methods use uniform settings and the same optimization loss on the benchmarks. Appendix B introduces the detailed settings of hyperparameters. Remarkably, CryptoGAT outperforms all state-of-the-art (SOTA) models and far surpasses other models in the Sharpe ratio, a key investment metric. LSTM and GRU show moderate predictive power. Conversely, ALSTM and Transformer architectures, which have more complex time modeling, performed worse. Similarly, StockMixer -- the current state-of-the-art for stock prediction -- achieves an IC of only 0.011 with a low Sharpe (0.356) among most models. Another interesting observation is that Spatio-temporal models that augment a temporal backbone with graph modules outperform complex pure time series models. However, because they focus more on the complexity of the time dimension, their model performance and inference efficiency are far inferior to our graph models.

Overall, CryptoGAT achieved the best overall performance while maintaining competitive inference speed. Compared to the best spatio-temporal baseline (MASTER, IC\,=\,0.027), FGAT improves IC by 74\% with only one-third of the parameters. This also demonstrates the potential of redefining the pure price prediction task of cryptocurrencies as a graph problem.

\begin{table}[t]
\centering
\caption{Portfolio performance of CryptoGAT under no-cost and realistic transaction costs.}
\label{tab:portfolio_performance}
\setlength{\tabcolsep}{0pt}
\small
\begin{tabular*}{\dimexpr\columnwidth-2em\relax}{@{\extracolsep{\fill}}lcc@{}}
\toprule
\textbf{Metric} & \textbf{No Cost} & \textbf{Cost(10 bps/side)} \\
\midrule
IC                  & 0.037            & 0.037            \\
Cumulative Return   & +229.12\% & +185.66\%         \\
Annualized Return   & +779.37\% & +579.10\%         \\
Volatility/day      & 4.25\%           & 4.25\%           \\
Sharpe (ann.)       & \textbf{3.128}    & \textbf{2.759}             \\
Max Drawdown        & -36.63\%        & -38.98\%        \\
Avg Turnover/day    & 35.5\%            & 35.5\%            \\
\bottomrule
\end{tabular*}
\end{table}

\subsection{Comprehensive Portfolio Evaluation}

To ensure that our findings can be applied to real-world investment scenarios, we further validate our model through a comprehensive portfolio evaluation metric, including cumulative returns, volatility, Sharpe ratio, maximum drawdown, and turnover rate, and conduct a transaction cost sensitivity analysis. 

Table~\ref{tab:portfolio_performance} reports the portfolio performance of GAT under two scenarios: No Cost and Cost, 10 bps per side. The model delivers a Sharpe ratio of 3.128 and a cumulative return of 229.12\%, demonstrating strong resilience even under a maximum drawdown of -36.63\%, and confirms the model sustains consistent positive returns with downside risk. The daily volatility of 4.25\% and the average daily turnover rate of 35.5\% demonstrate that the model's returns stem from its trading capabilities. Furthermore, under actual transaction costs, the Sharpe ratio only slightly decreased from 3.128 to 2.759, indicating the model's stability in real-world trading environments.

To verify that the profitability is not contingent on a specific cost assumption, we further provide a cost sensitivity analysis based on cost calculations \cite{makarov2020trading}. Table~\ref{tab:cost_sensitivity} reports a sensitivity analysis of the Sharpe ratio and annualized return across transaction costs ranging from 0 to 30 bps per side. Even at 30 bps per side (including slippage and trading fees), the model maintains a Sharpe Ratio above 2.0 and annualized returns exceeding 300\%, demonstrating its robustness with transaction-cost assumptions.

\begin{table}[t]
\centering
\caption{Cost sensitivity analysis of CryptoGAT.}
\label{tab:cost_sensitivity}
\setlength{\tabcolsep}{0pt}
\small
\begin{tabular*}{\dimexpr\columnwidth-2em\relax}{@{\extracolsep{\fill}}lcc@{}}
\toprule
\textbf{Cost (per side)} & \textbf{Ann. SR} & \textbf{Ann. Return} \\
\midrule
0.000\%               & \textbf{3.128}  & \textbf{+779\%}  \\
0.050\%               & 2.920           & +673\%           \\
0.075\%               & 2.840           & +624\%           \\
\textbf{0.100\%}      & \textbf{2.759}  & \textbf{+579\%}  \\
0.200\%               & 2.438           & +424\%           \\
0.300\%               & \textbf{2.117}           & \textbf{+305\%}           \\
\bottomrule
\end{tabular*}
\end{table}

\section{Empirical Analysis and Discovery}

\subsection{\textit{Ablation Study}}
\textbf{\textit{How effective are different components?}}

We validated the effectiveness of each of the three mixing modules in StockMixer by removing one of them individually, and compared it with our GAT model and a GAT model with added time mixing functionality. Comparative experiments were implemented on Nasdaq \cite{fan2024stockmixer} and cryptocurrency datasets. The Nasdaq benchmark dataset we used has been adopted by high-quality conferences, and the data processing is consistent with StockMixer \cite{fan2024stockmixer}. The results are shown in the table~\ref{tab:ablation_study}. In stock prediction, the time mixing module of StockMixer is the most important. After removing the time mixing module, the IC on Nasdaq decreased from 0.043 to 0.018. This demonstrates that learning effective temporal representations is crucial in stock prediction. At the same time, it also explains why StockMixer and previous stock prediction architectures have adopted a time-first framework. 

However, in cryptocurrency prediction tasks, the time mixing module and StockMixer exhibit substantially inferior performance compared to the GAT model, which relies solely on graph structure modeling. To further validate our findings, we fused the time mixing module with GAT and conducted comparative experiments. Remarkably, introducing the time mixing module not only failed to improve model performance but also weakened GAT's predictive ability: IC dropped from 0.037 to 0.006, and ICIR from 0.138 to 0.023. This further demonstrates the paper’s hypothesis.

\begin{table}[t]
\centering
\caption{Ablation study of model components on NASDAQ and
Cryptocurrency.}
\label{tab:ablation_study}
\setlength{\tabcolsep}{0pt}
\small
\begin{tabular*}{\dimexpr\columnwidth-2em\relax}{@{\extracolsep{\fill}}ccccc@{}}
\toprule
\textbf{Model Component} & \multicolumn{2}{c}{\textbf{NASDAQ}} & \multicolumn{2}{c}{\textbf{Cryptocurrency}} \\
\cmidrule(lr){2-3} \cmidrule(lr){4-5}
 & \textbf{IC} & \textbf{ICIR} & \textbf{IC} & \textbf{ICIR} \\
\midrule
\textbf{StockMixer} & \textbf{0.043} & \textbf{0.501} & 0.011 & 0.113 \\
w.o. Indicator Mix. & 0.040 & 0.465 & -0.008 & -0.052 \\
w.o. Time Mix. & 0.018 & 0.164 & -0.014 & -0.071 \\
w.o. Stock Mix. & 0.037 & 0.376 & 0.001 & 0.011 \\
\textbf{GAT} & 0.035 & 0.377 & \textbf{0.037} & \textbf{0.138} \\
GAT + Time Mixing  & -0.008  & -0.117 & 0.006 & 0.023 \\
\bottomrule
\end{tabular*}
\end{table}

\subsection{Empirical Analysis on Input Sequences}
\textbf{\textit{Can time series models learn useful information about cryptocurrencies from longer input sequences?}}

The size of the look-back window has a significant impact on prediction accuracy because it directly determines how much information the model can acquire during training. Generally, models with effective information extraction capabilities should be able to achieve better results using larger look-back window sizes. To systematically evaluate whether time series models can extract effective predictive signals from cryptocurrency data, we set different look-back window sizes $L \in \{5, 10, 15, 30, 60\}$. At the same time, we compared our GAT model with the four most representative or high-performing time series models from different categories.

\begin{figure}[t]
  \centering
  \includegraphics[width=\dimexpr\linewidth-2em\relax]{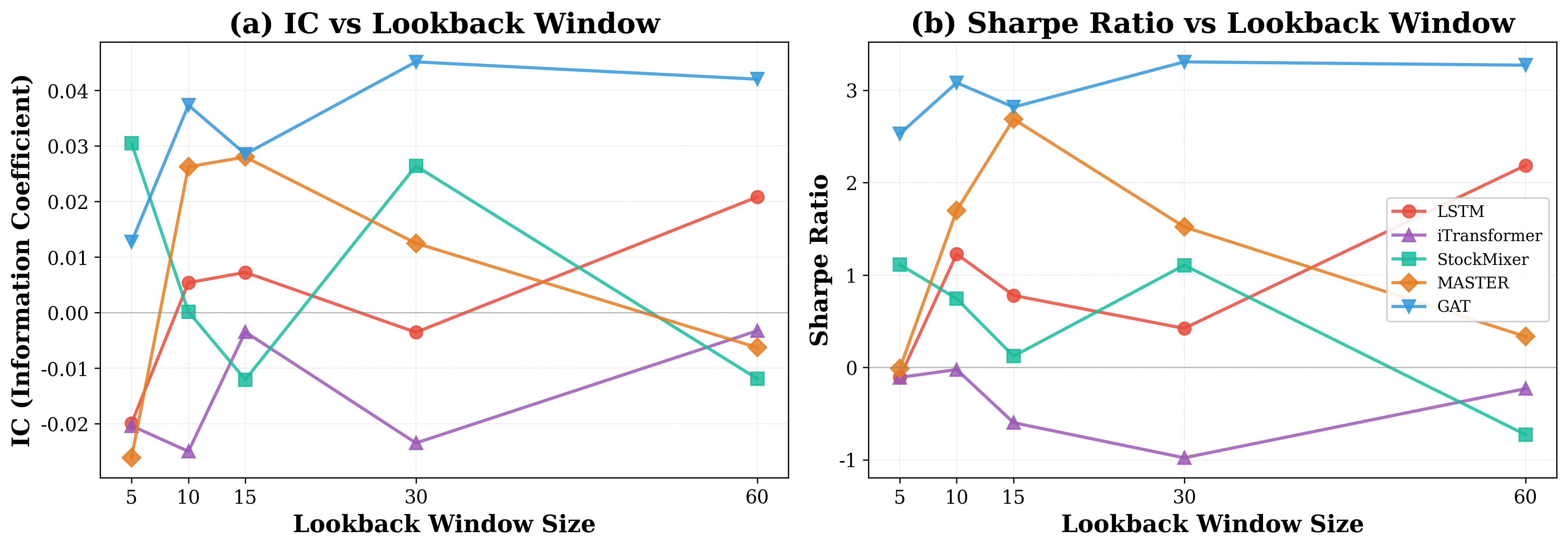}
  \setlength{\abovecaptionskip}{2pt}
  \caption{IC and annualized Sharpe ratio (Y-axis) across different lookback window sizes (X-axis).}
  \label{fig:ic_lookback}
\end{figure}

Figure~\ref{fig:ic_lookback} shows the Information Coefficient (IC) and Sharpe ratio results on the cryptocurrency dataset. Consistent with our expectations, the performance of most time-series-based models decreases with increasing look-back window size. The IC and Sharpe ratio of iTransformer remain negative, indicating that this powerful time series model does not adapt well. In contrast, the performance of the GAT (Graph Attention Network) model improves with increasing look-back window size and gradually stabilizes. This further confirms our hypothesis that a lightweight graph model consistently and significantly outperforms complex state-of-the-art temporal models, underscoring the importance of spatial structure over temporal complexity in cryptocurrency.

\subsection{\textit{Empirical Analysis on Datasets}}
\textbf{\textit{Why Do Time Series Models Underperform on Cryptocurrency Data?}}

The numerous experiments above have demonstrated that time series models struggle to learn effective information from cryptocurrency datasets. Why do time series models perform well on stock data but underperform on cryptocurrency datasets? This performance gap persists across different look-back windows and model architectures, indicating that the root cause lies not in these factors, but in a fundamental difference in the data itself. To reveal this reason, we conducted a systematic analysis of the internal characteristics of stocks and cryptocurrency data.

\textbf{1. Predictability of the Signal}

To fairly quantify the predictability of the signal, we set up autoregressive (AR) models with different lookback windows $L \in \{5, 10, 20, 30\}$. The AR model is a simple and clean benchmark that measures time dependence. We used an AR model as a measurement tool to evaluate the amount of predictable information contained in 1,026 Nasdaq stocks \cite{fan2024stockmixer} and cryptocurrencies along the temporal dimension and used the coefficient of determination $R^2$ to measure predictability, as follows:

\begin{equation}
    R^2 = 1 - \frac{\sum_{i=1}^{n} (y_i - \hat{y}_i)^2}{\sum_{i=1}^{n} (y_i - \bar{y})^2}
\end{equation}

\noindent where $y_i$ is the actual value, $\hat{y}_i$ is the predicted value, $\bar{y}$ represents the mean value of all actual samples. The coefficient of determination $R^2$ measures the predictive ability of historical signals for future values, directly reflecting the predictability of the dataset. where a higher $R^2$ indicates stronger predictability. 

The standard deviation curve in Figure 3(a) shows that the R² standard deviation of Nasdaq is more than 5 times that of cryptocurrencies. This means that in time-dimension forecasting, the predictability of Nasdaq stock data is far superior to that of cryptocurrency data. Furthermore, within the framework of time-series AR model forecasting experiments, the predictability of stock data gradually improves with increasing lookback windows, while the predictability of cryptocurrency data nearly stagnates. For stock data, a larger lookback window (more historical information), in the time dimension, leads to better performance and a higher predictability $R^2$. However, for cryptocurrency data, increasing the lookback window, in the time dimension, has little impact on model performance and predictability $R^2$. Meanwhile, the distribution of R² also indicates that the predictability of cryptocurrency assets is consistently below 0.1 (see Figure 3(b)), while the predictability of most stock assets is much higher. This further demonstrates that cryptocurrency data lacks sufficient and effective time-dimensional information for time series models to learn from. This also explains why time series models often underperform in cryptocurrency prediction tasks.

\begin{figure}[t]
  \centering
  \includegraphics[width=\dimexpr\linewidth-2em\relax]{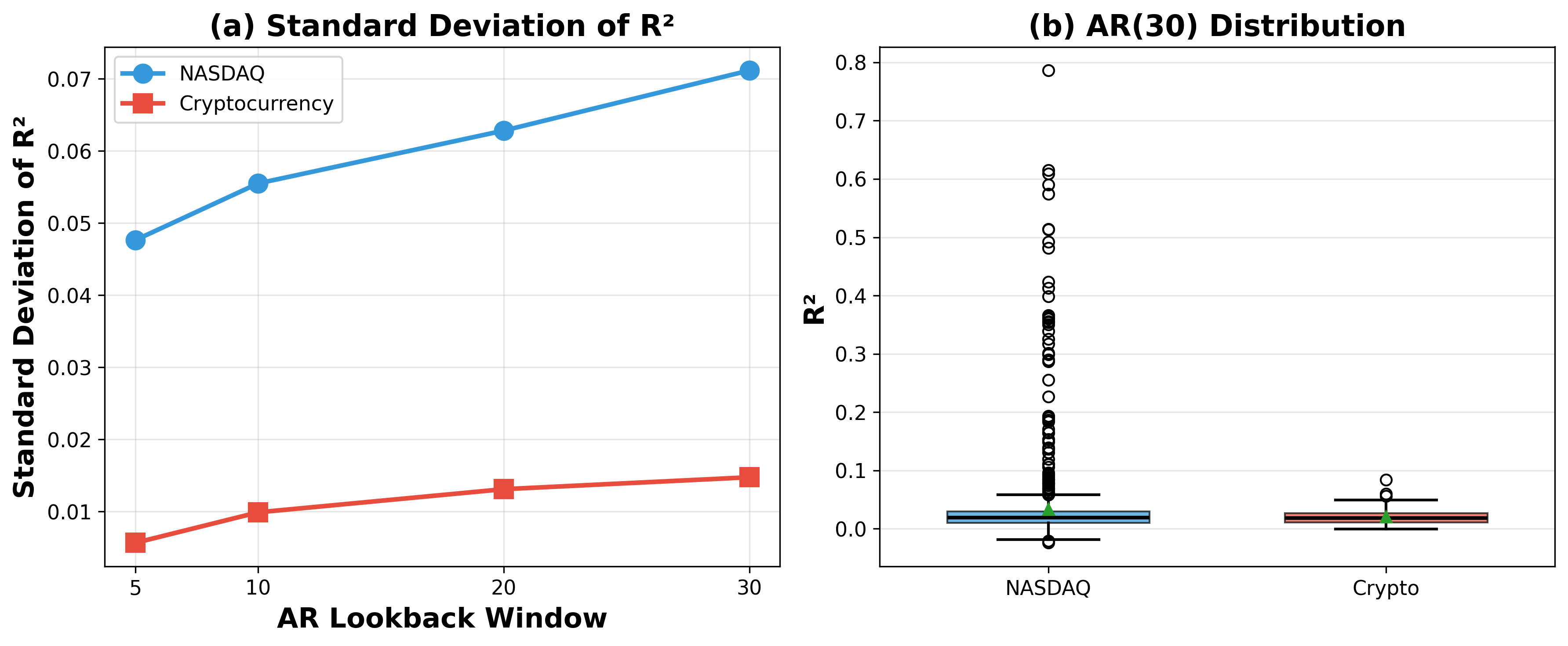}
  \setlength{\abovecaptionskip}{2pt}
  \caption{The standard deviation and distribution of $R^2$ (Y-axis) of models
with different look-back window sizes (X-axis).}
\setlength{\abovecaptionskip}{2pt}
\end{figure}

\textbf{2. Cross-Asset Dependencies}

To gain a deeper understanding of the inherent differences between Nasdaq and cryptocurrency data, we constructed an evaluation system based on three dimensions: Statistical Correlation Strength, Graph Density Metrics, and Principal Component Analysis (PCA). 

\textbf{2.1) Statistical Correlation Strength} is used to intuitively assess consistency among assets. It measures the correlation between two variables using the Pearson correlation coefficient $\rho_{X,Y}$. The formula is as follows: $\rho_{X,Y} = \frac{\text{cov}(X,Y)}{\sigma_X \sigma_Y} = \frac{\sum_{i=1}^{n} (X_i - \bar{X})(Y_i - \bar{Y})}{\sqrt{\sum_{i=1}^{n} (X_i - \bar{X})^2} \sqrt{\sum_{i=1}^{n} (Y_i - \bar{Y})^2}}$, The correlation coefficient is obtained by calculating the covariance of the two variables $X_i, Y_i$ relative to their means $\bar{X}, \bar{Y}$ and then dividing it by the product of their standard deviations. Its value ranges from -1 to 1. The closer the value is to 1, the stronger the correlation between the data.

As shown in Figure 4(a), the average correlation coefficient of cryptocurrencies (0.528) is significantly higher than that of stocks (0.200). Notably, 65.78\% of cryptocurrency trading pairs exhibit strong correlation (r $>$ 0.5), compared to only 2.39\% of stock trading pairs. The analysis reveals that cryptocurrency assets exhibit strong cross-asset relationships, which is precisely what graph-based architectures are designed to capture.

\begin{figure}[h]
  \centering
  \includegraphics[width=\dimexpr\linewidth-2em\relax]{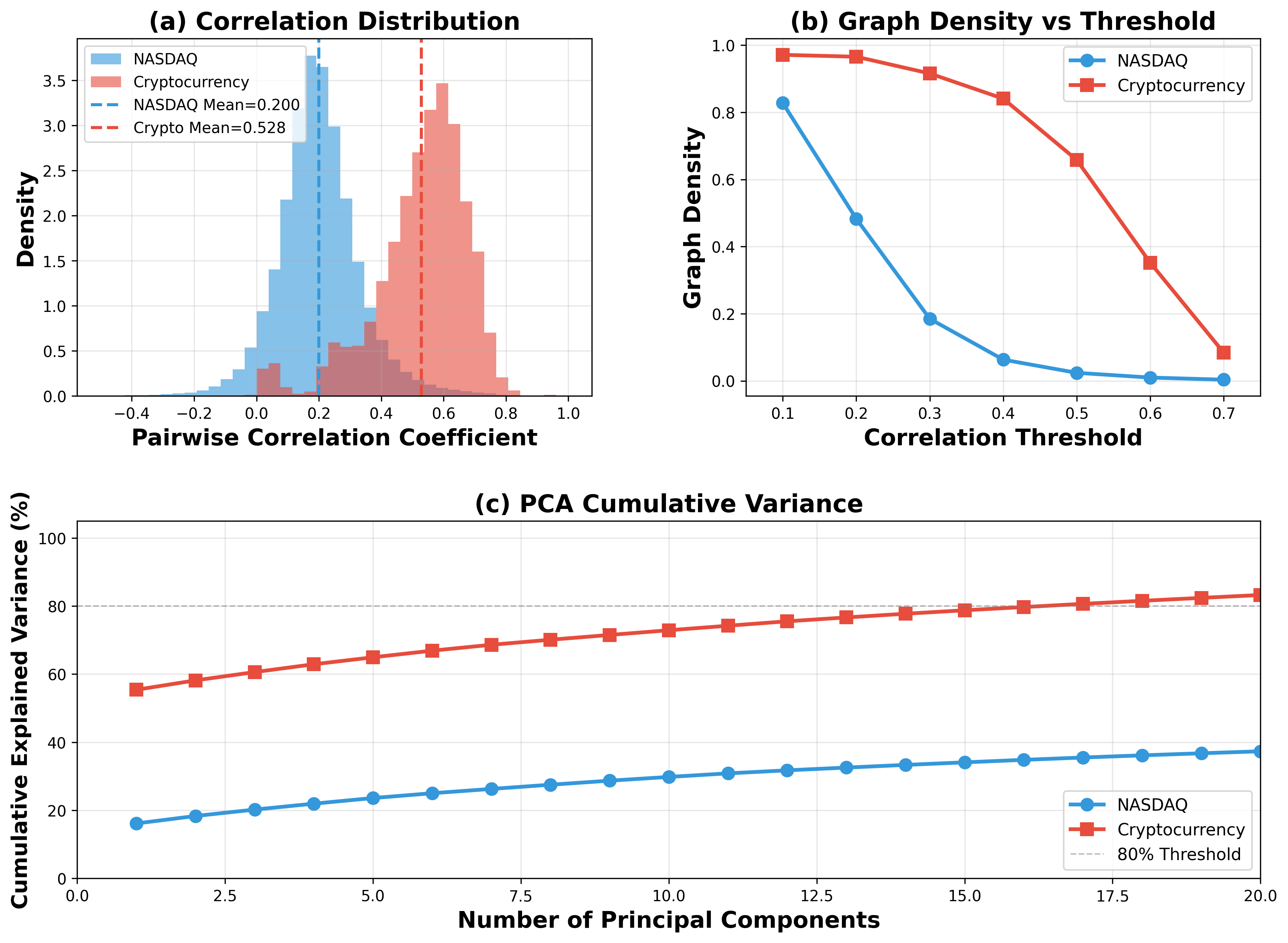}
  \setlength{\abovecaptionskip}{2pt}
  \caption{The comparison of Correlation Strength, Graph Topological, and Principal Component Analysis (PCA).}
\end{figure}

\textbf{2.2) Graph Density} assesses the tightness of the data structure under varying thresholds. We model the market as an undirected graph $G(N,E)$, where nodes $N$ are assets and an edge $(i,j)\in E$ exists whenever the absolute pairwise correlation $|\rho_{ij}|$ exceeds a threshold $\tau$. Under this construction, the graph density $D(\tau) = \frac{2|E(\tau)|}{N(N-1)}$ measures the fraction of asset pairs whose co-movement is strong enough to survive the cutoff $\tau$. Sweeping $\tau$ from low to high progressively strips away weaker, noisier links, so the curve $D(\tau)$ directly traces how strong the typical pairwise relationship in a market actually is. If density remains high under a very strict correlation threshold (tending towards 1), it indicates that a large number of assets in the market are closely connected.

Figure 4(b) shows that the tightness of the Nasdaq data structure decreases rapidly as the correlation threshold increases. When the threshold is set to 0.5, its tightness has dropped from an initial 0.8 to 0.2. Cryptocurrencies, however, maintain a tightness as high as 0.66 at this point. The stark contrast reveals that cryptocurrency markets exhibit spatial dependence far stronger than equity markets. This persistent, high-strength coupling provides the structural signal that graph-based models can directly exploit, yet purely temporal models struggle to capture.

\textbf{2.3) Principal Component Analysis (PCA)} is used to measure the most prominent hidden patterns of change in data. It is a data dimensionality reduction technique that extracts the main information (variance) of the data by transforming the original data. We therefore use the variance share of PC1 as a probe of market homogeneity, the higher the share, the more tightly the assets move along a single common pattern. In financial markets, its leading component has a clear interpretation: it is the market factor shared by all assets. Given centered data $X \in \mathbb{R}^{n\times f}$ with $n$ samples and $f$ features, we form the covariance matrix $C=\frac{1}{n-1}X^{\top}X,$ perform the eigenvalue decomposition $C v_k = \lambda_k v_k$ (eigenvectors $v_k$ and eigenvalues $\lambda_k$), and compute the variance ratio of the $k$-th component as $\text{Ratio}_k = \frac{\lambda_k}{\sum_{j=1}^{n} \lambda_j}$ ($\sum_{j=1}^{n} \lambda_j$ is the total variance of all features). $\text{Ratio}_k$ is the percentage of variance explained by the k-th principal component relative to the total variance. The first component PC1, corresponding to the largest eigenvalue $\lambda_1$, captures the single linear direction along which the data vary most. A large $\text{Ratio}_1$ thus directly quantifies how much of the total market dynamics is driven by one shared pattern.

The first principal component $PC_1$ in the cryptocurrency market accounts for a staggering 55\%, while the proportion in the Nasdaq stock market is only 16\% (see Figure 4(c)). When evaluating the top 20 principal components, the cumulative component proportion of cryptocurrency data exceeds 80\%, higher than that of stock data. This high co-movement concentration indicates that cryptocurrency assets are tightly interconnected, forming a natural graph structure where graph-based modeling can effectively exploit these dense inter-asset relationships.

\textbf{\textit{2.3.1) Does the Market Dominant Factor Replace Graph Structure?}}

The high PC1 (55\%) can easily lead to a common misconception: whether the cryptocurrency market is influenced by the market-wide factors. If so, a linear model that includes the market-dominant factors might achieve similar performance. To further validate the importance of graph structure, we conducted a more comprehensive comparative analysis. We added market-dominant return factors as the sixth feature to the input of each asset and retrained four of the most representative or highest-performing time series models from different categories under the same training protocol.

\begin{table}[t]
\centering
\caption{All models share the identical training protocol.}
\label{tab:mf-comparison}
\setlength{\tabcolsep}{0pt}
\small
\begin{tabular*}{\dimexpr\columnwidth-2em\relax}{@{\extracolsep{\fill}}lccc@{}}
\toprule
\textbf{Model} & \textbf{IC} & \textbf{ICIR} & \textbf{Sharpe} \\
\midrule
MF-LSTM          & 0.009           & 0.048           & 0.669 \\
MF-iTransformer  & -0.012        & -0.091        & -0.591 \\
MF-StockMixer    & 0.015           & 0.094  & 0.963 \\
MF-MASTER        & 0.008           & 0.027           & 0.926 \\
\textbf{GAT}  & \textbf{0.037}  & \textbf{0.138}           & \textbf{3.128} \\
\bottomrule
\end{tabular*}
\end{table}

Table~\ref{tab:mf-comparison} shows that the market-dominant factor is insufficient for effective prediction. The four temporal baselines with market-dominant factors underperform our GAT. This also clarifies the common misconception: the model's performance does not stem from the market-dominant factor, but rather from the graph model's learning of heterogeneous pairwise relationships.

\section{Conclusion and Future Work}

\textbf{Conclusion.}  To our knowledge, due to bias in problem definition, pure cryptocurrency price prediction models have rarely appeared in high-quality conferences. This bias in problem definition has driven the field research toward incorporating external and multimodal signals (such as on-chain data and sentiment features) to compensate for the limited predictive power. Our work takes the opposite approach: rather than augmenting data sources, we \textbf{recast pure price-based cryptocurrency prediction from a time-series problem into a graph problem}.

Through the extensive and comprehensive experimental analysis, we demonstrate that cryptocurrency data is a graph structure with rich heterogeneous pairwise relationships, rather than simple time series data. We use a set of simple yet effective graph attention models, CryptoGAT, to validate our argument. Extensive experiments on real-world cryptocurrency datasets demonstrate that our proposed CryptoGAT models outperform various state-of-the-art prediction methods. It is important to note that our contribution is not merely to propose a graph attention model, but rather to recast an important question, present comprehensive experimental analysis, and demonstrate from multiple perspectives the effectiveness of graph-based modeling. Our comprehensive research will contribute to future research in this field.

\textbf{Future work.}  Our redefinition and analysis of the cryptocurrency prediction task can help researchers in this field open up new research perspectives, provide a reproducible foundation for graph-centric modeling of cryptocurrency, and inspire research enthusiasm in new model design and data processing. CryptoGAT, with its strong interpretability, serves as simple yet competitive benchmark models that can be used and compared in future research.


\appendix

\section{DETAILS ON EXPERIMENTAL SETUP}

\subsection{Baseline methods}

To rigorously evaluate the efficacy of our proposed architecture, we benchmark it against a diverse set of state-of-the-art methods, ranging from fundamental sequence models to advanced Spatio-Temporal models.

\textbf{Classical RNN Models:}

\begin{itemize}
    \item \textbf{LSTM.} \cite{hochreiter1997long} It is a recurrent neural network that employs the standard Long Short-Term Memory network to process temporal price sequences.
    \item \textbf{ALSTM.} \cite{feng2019enhancing} It is an enhanced recurrent framework that incorporates adversarial training and stochastic simulation into the LSTM structure, aimed at capturing the stochastic nature of market dynamics more robustly.
    \item \textbf{GRU.} \cite{cho2014learning} The Gated Recurrent Unit (GRU) is a variant of recurrent neural networks that addresses the vanishing gradient problem present in traditional RNNs. It simplifies the LSTM architecture by merging the forget gate and input gate into a single update gate, thus more efficiently capturing temporal dependencies. 
\end{itemize}

\textbf{Transformer Models:}

\begin{itemize}
    \item \textbf{PatchTST.} \cite{nie2023time} It segments time series into patched subsequences and applies a channel-independent Transformer encoder for multivariate forecasting.
    \item \textbf{iTransformer.} \cite{liu2024itransformer} It reverses the dimensionality processing logic of Transformer, treating each time step as a whole feature vector and applying a self-attention mechanism on the feature dimension rather than the time dimension. This design subverts the traditional Transformer architecture.
\end{itemize}

\textbf{GNNs Models:}

\begin{itemize}
    \item \textbf{RSR.} \cite{feng2019temporal} It is a hybrid architecture fusing Temporal Graph Convolution with LSTM to capture time-sensitive stock interactions. We adopt the \textbf{RSR-I} using neural network variant as the baseline, as it empirically outperforms the RSR-E using similarity variant.
    \item \textbf{ESTIMATE.} \cite{huynh2023efficient} It is a memory-augmented LSTM architecture designed to capture individual asset patterns. Uniquely, it employs hypergraph attention to model non-pairwise correlations.
\end{itemize}

\textbf{Spatio-Temporal Models:}

\begin{itemize}

    \item \textbf{TGN.} \cite{rossi2020temporal} It is a memory-based framework for continuous-time dynamic graphs, where a sequential per-node memory processes time-stamped events as the primary mechanism and graph message passing is layered on top to route these temporal updates between neighboring nodes.

    \item \textbf{Graphormer.} \cite{ying2021transformers} It is a graph transformer that encodes graph topology through centrality, spatial, and edge biases in self-attention; originally proposed for static graph-level tasks, its adaptation to spatio-temporal forecasting commonly applies the graph-attention block per time step within a broader temporal pipeline.
    
    \item \textbf{MASTER.} \cite{li2024master} It is a market-guided stock transformer whose core mechanism is temporal self-attention over each asset's own sequence, with cross-asset attention introduced as a secondary stage in which an aggregate market token conditions inter-stock interactions.

\end{itemize}

\textbf{SOTA Models:}

\begin{itemize}
    \item \textbf{StockMixer.} \cite{fan2024stockmixer} It is a stock time series mixing model based on MLP-Mixer, which consists of three parts: indicator mixing, time mixing, and stock mixing.
\end{itemize}

\subsection{Hyper-parameter settings}

\subsubsection{Parameter settings}

To ensure the fairness and consistency of the experiment, we used a uniform 30-day lookback window length and consistently set the learning rate to 0.00025 and Dropout to 0 throughout the 100 training epochs. We conduct experiments using the following parameter settings 
as shown in Table~\ref{tab:parameters}.

\begin{table}[h]
\centering
\caption{Default parameter settings}
\label{tab:parameters}
\begin{tabular*}{\dimexpr\columnwidth-4em\relax}{@{\extracolsep{\fill}}lc@{}}
\toprule
\textbf{Parameter} & \textbf{Values} \\
\midrule
Lookback window length  & 30 \\
Learning rate & 0.00025 \\
Train, validate, test split & 60\%,20\%,20\% \\
Dropout & 0 \\
Training epochs & 100 \\
Hidden size & 64 \\
\bottomrule
\end{tabular*}
\end{table}

\end{document}